\documentclass[aip,preprint,numerical,a4paper,showpacs]{revtex4-1}

\usepackage{graphicx}
\usepackage{amsmath}
\usepackage{bbm}
\usepackage{amsfonts}
\usepackage{amssymb}
\usepackage{amsbsy}
\usepackage{dsfont}
\usepackage[T1]{fontenc}
\usepackage{theorem}
\usepackage{stmaryrd}
\usepackage{tipa}
\usepackage{textcomp}
\usepackage{subfigure}
\usepackage{epsfig}
\usepackage{lmodern}
\usepackage{framed}
\usepackage[subfigure]{tocloft}
\usepackage{subeqnarray}
\usepackage{alphalph}
\usepackage[refpage,cfg,prefix]{nomencl}
\usepackage{multirow}
\usepackage{xifthen}
\usepackage{enumerate}
\usepackage{color}
\usepackage{wasysym}
\usepackage{bibentry}
\usepackage{mathtools}
\usepackage{amssymb}
\usepackage{mhchem}
\nobibliography*
\usepackage{verbatim}
\usepackage{url}
\usepackage{setspace}
\usepackage{xr}
\usepackage{diagbox}

\makenomenclature

\DeclarePairedDelimiter\floor{\lfloor}{\rfloor}
\DeclareMathOperator{\erf}{erf}

\DeclareMathOperator{\sgn}{sgn}

\begin{document}
\title[A hybrid algorithm for coupling PDE and compartment-based dynamics]{A hybrid algorithm for coupling PDE and compartment-based dynamics}
\author{Jonathan U. Harrison}
\email[]{harrison@maths.ox.ac.uk}
\affiliation{Wolfson Centre for Mathematical Biology, Mathematical Institute, University of Oxford, Andrew Wiles Building, Radcliffe Observatory Quarter, Woodstock Road, Oxford, OX2 6GG, United Kingdom}
\author{Christian A. Yates}
\email[]{c.yates@bath.ac.uk}
\affiliation{Department of Mathematical Sciences, University of Bath, Claverton Down, Bath, BA2 7AY, United Kingdom}
\homepage{Website: http://people.bath.ac.uk/cy386/}
\date{\today}

\begin{abstract}
Stochastic simulation methods can be applied successfully to model exact spatio-temporally resolved reaction-diffusion systems.
However, in many cases, these methods can quickly become extremely computationally intensive with increasing particle numbers.
An alternative description of many of these systems can be derived in the diffusive limit as a deterministic, continuum system of partial differential equations.
Although the numerical solution of such partial differential equations is, in general, much more efficient than the full stochastic simulation, the deterministic continuum description is generally not valid when copy numbers are low and stochastic effects dominate.

Therefore, to take advantage of the benefits of both of these types of models, each of which may be appropriate in different parts of a spatial domain, we have developed an algorithm that can be used to couple these two types of model together.
This hybrid coupling algorithm uses an overlap region between the two modelling regimes.
By coupling fluxes at one end of the interface and using a concentration-matching condition at the other end, we ensure that mass is appropriately transferred between PDE- and compartment-based regimes. 
Our methodology gives notable reductions in simulation time in comparison with using a fully stochastic model, whilst maintaining the important stochastic features of the system and providing detail in appropriate areas of the domain.
We test our hybrid methodology robustly by applying it to several biologically motivated problems including diffusion and morphogen gradient formation. 
Our analysis shows that the resulting error is small, unbiased and does not grow over time.
%
\end{abstract}

\maketitle

\section*{}\label{introduction}

Multiscale modelling challenges occur frequently throughout cellular biology and in the context of cell migration.
Spatial reaction-diffusion models can be used to describe, either deterministically or stochastically, various biological phenomena.
These include actin dynamics in filopodia \citep{erban2013multiscale}, calcium signalling \citep{rudiger2007hybrid}  and chemisorption of polymers \citep{franz2012multiscale}.
In many cases, it may be beneficial to use a multiscale approach to modelling using different descriptions in different spatial regions.
In this article, we will set out a method for coupling together a continuum deterministic description and a discrete stochastic description. 

Commonly, continuum approaches using partial differential equations (PDEs) are adopted to model biological systems \citep{keller1971traveling}.
These equations can either be solved analytically (in some cases) or simulated numerically. 
Results using this methodology are relatively fast to calculate computationally.
However, for systems with small numbers of molecules the results obtained using deterministic methods may not always capture the behaviour of a stochastic system appropriately, especially in situations where molecular numbers are low and interactions are non-linear. 
For example, in a system with multiple steady states such as the canonical model of \citet{Schloegl1972chemical}, a deterministic model fails to capture the switching behaviour between the steady states seen in a stochastic model.
In general, PDE models break down when the number of molecules present is very low and stochastic effects dominate \citep{franz2013hybrid}. 
Although deterministic models may provide useful information about average behaviour (in the case of linear systems) they cannot offer a full description of every system.
Thus, in the case where copy numbers are low, the best description will be afforded by a stochastic model.
There are two main types of stochastic models used for reaction-diffusion equations \citep{flegg2012two}: off-lattice methods and on-lattice compartment-based methods.
We will focus on compartment-based methods, which generally offer a coarser description than their off-lattice counterparts.

When simulating a system using a compartment-based stochastic model (also known as a position-jump model), the computational cost of the simulations can become prohibitive if the number of particles in the system is high. 
A computationally efficient continuum model may be more appropriate in this scenario. 
Thus in situations where particle concentrations vary widely across the domain there may be advantages to using a continuum PDE model in the region of the spatial domain where particle numbers are high and a discrete stochastic model elsewhere.
Moreover, detail is often only required in a certain part of the domain and thus a spatial-hybrid model may be most appropriate \citep{franz2012multiscale, erban2013multiscale, flegg2012two, flekkoy2001coupling, moro2004hybrid, ferm2010adaptive}.
Such a hybrid model would allow an accurate representation of the reaction-diffusion dynamics in the region where this is required but minimises the computational resources needed to perform the calculation by using less detailed, more efficient methods in regions of the domain where detail is not required.


Previously, \citet{flekkoy2001coupling} have developed a hybrid model that links a PDE-based model to the motion of random walkers on a lattice.
Motivated by heat transport around a facture in a solid, \citet{flekkoy2001coupling} choose a detailed description of the particle dynamics coupled to a coarse-grained PDE model:
the lattice spacing used to solve the PDE is larger than in the particle-based region. 
More recently, PDE-to-compartment hybrid methods have been developed which employ a region of the PDE regime in which particles are represented using both the compartment- and PDE-based modelling regimes simultaneously \citep{yates2015pseudo,spill2015hybrid}. 
The duality of these so-called ``pseudo-compartment methods'' allows for particles to behave correctly as they cross individually between the two different regimes since particles can jump into their neighbouring compartment according to standard compartment-based rules for diffusion.

Our hybrid modelling regime employs a PDE mesh that is significantly (and arbitrarily) finer than the lattice in the compartment-based region.
This choice is natural in many situations, including in a biological context, where we are choosing to use the PDE model in regions of high population to offer improved computational efficiency.
Taking a fine mesh will not prove computationally prohibitive compared to the stochastic model, but allows us to make the numerical solution of the PDE arbitrarily accurate. Methodologies with coarser or equal PDE spacing relative to compartment spacing \citep{flekkoy2001coupling,spill2015hybrid} are open to questions about what exactly the ``PDE regime'' represents given its resolution and accuracy are restricted by the resolution of the compartment-based method.

Our approach to coupling of deterministic PDE-based and stochastic compartment-based regions employs an overlap region where both modelling descriptions are valid.
This overlap region can contain multiple compartments if desired.
The method that we have developed relies upon specifying a Dirichlet-type condition between the two models at one interface at the edge of the PDE-based region and dictating the correct flux of PDE on compartments at the other interface.
This fixes the boundary conditions at the interfaces between each of the regions.

In the remainder of this article, we describe and explore our novel hybrid coupling algorithm in detail and illustrate the effectiveness of the method. 
In section \ref{section:coupling_algorithm}, we present the hybrid method in full and justify the coupling conditions chosen.
Thereafter, in section \ref{section:numerical_simulations}, we demonstrate the appropriate behaviour of our method through its application to systems of diffusing particles with various extreme initial conditions (chosen specifically to test the algorithm) and a biologically-motivated example: the formation of a morphogen gradient.
We apply the model to a travelling wave example in section \ref{section:travelling} and introduce an adaptive interface between the modelling regimes in section \ref{section:adaptive}.
We present detailed simulation-time comparisons of the hybrid model with the fully stochastic model for our test problems at the end of section \ref{section:adaptive}, which explicitly demonstrate the improved efficiency of our hybrid method.
The fidelity of the algorithm's performance is then examined and the error (with respect to a range of model parameters) analysed in section \ref{section:error_analysis}.
We conclude in section \ref{section:discussion}, with a discussion of the potential advantages of this hybrid method in relation to other existing methods.

\section{Methods}
\subsection{The domain} \label{section:domain}

\begin{figure}[h!]
\centering
\includegraphics[width=0.8\columnwidth]{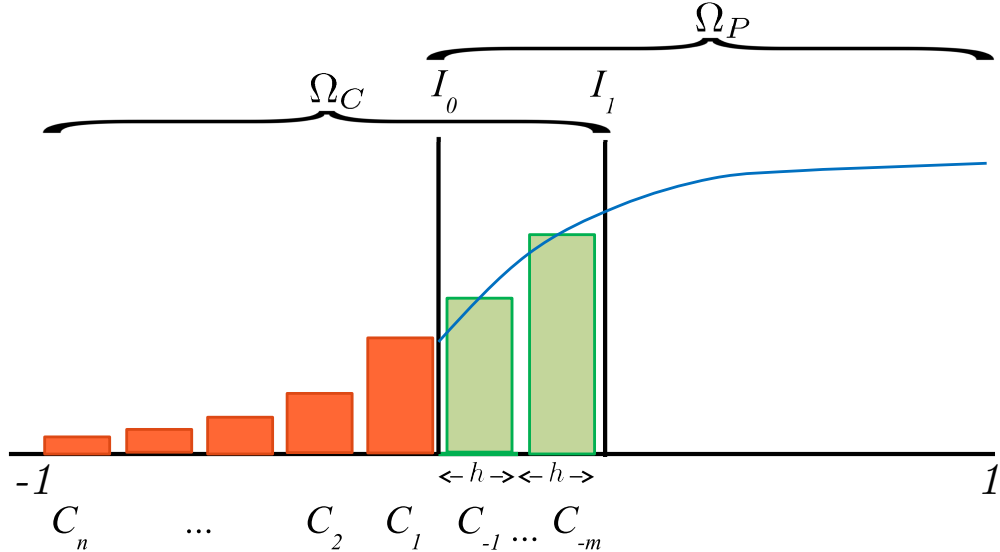}
\caption[The domain $\Omega$.]{The domain $\Omega $ showing the division into a compartment-based region $\Omega_c$ on $[-1,I_1]$ and a PDE-based region $\Omega_p$ on $[I_0,1]$ with an overlap region where both model descriptions are valid on $[I_0,I_1]$. 
Orange bars represent the number of particles in the fully compartment-based regime, green bars represent the number of particles in each compartment of the overlap region, and the blue curve represents the solution of the PDE.}
\label{figDomain}
\end{figure}

Suppose, arbitrarily, we have a domain $\Omega = [-1,1]$ which we divide into a region $\Omega_c$ in which we use a compartment-based, stochastic model and a region $\Omega_p$ in which we use a deterministic, PDE-based model.
A characterising feature of our hybrid methodology is an overlap region (shown in Figure \ref{figDomain}) in which both modelling regimes are simultaneously valid descriptions (i.e. $\Omega_c\cap\Omega_p\neq\emptyset$).
Either side of the overlap region, we have interfaces $I_0$ and $I_1$ (see Figure \ref{figDomain}).
In a similar context, it has been demonstrated that an overlap region is required to give the appropriate variance for a model coupling a Brownian motion particle-based description and a PDE-based model \citep{franz2012multiscale}.

In $\Omega_c $, the domain is split into compartments of width $h$, where the $k$th compartment occupies the region $[I_0-kh, I_0-(k-1)h]$ for $k=-m,..,-1,1,...,n $. The $m$ compartments labelled $-m,\dots, -1$ are situated in the overlap region and the $n$ compartments\footnote{Note there is no compartment $0$.} labelled $1,\dots, n$ are in $[-1,I_0]$.
The labelling of compartments is illustrated in Figure \ref{figDomain}. 
We assume particles are well mixed within these compartments.

A continuum description of diffusion, as assumed when modelling with a PDE, requires sufficiently high particle numbers.
For low concentrations, this assumption breaks down.
If the concentration of particles is $u(x,t)$, given a total of $N$ particles in the system, then we can relate the probability of finding any particular particle in the system, $p_p(x,t)$, to the concentration as $p_p(x,t)=u(x,t)/N$.
This probability density remains well defined even at low particle copy numbers, when we cannot interpret the PDE as a concentration but are able to view it as a probability density.
Therefore we interpret $p_p(x,t)$ as the probability density to find each of $N$ particles in the PDE region at a given location $x$ at a certain time $t$. The expected number of particles in a subset, $\omega$, of the PDE domain, $\Omega_p$, is given by $N\int_\omega p_p(x,t) dx$. We will use $u_k(t)$ where $k=1,...,K+1$ to denote the PDE density at the $k$th PDE lattice point in the finite difference discretisation of the PDE required for our hybrid algorithm. 

For the compartment-based regime, let $p_c(x,t)$ (defined initially only at the centre of compartments) be the probability of finding one of the identically initialised particles at position $x$ at time $t$.
Since each compartment is well mixed, we can describe the evolution of $p_c(x,t)$ using the reaction-diffusion master equation \citep{gardiner1985handbook}.
We will also use the notation $\textbf{A(t)} = (A_{-m}(t),\allowbreak...,\allowbreak A_{-1}(t), A_1(t),\allowbreak ... , A_n(t))$ to represent the distribution of particle numbers across compartments.


\subsection{The coupling algorithm}\label{section:coupling_algorithm}

We now describe an algorithm which couples the two regimes together. Informally, the coupling is achieved by setting the value of the PDE lattice point at $I_0$ to the average of the adjacent compartment populations in $\Omega_c$ and using the gradient in the PDE-based region $\Omega_p$ to give a rate of jumping across the interface $I_1$ for the compartment-based regime.

In what follows, we specify and justify these coupling conditions mathematically.
These conditions are analogous a to Neumann condition for the compartments at $I_1$ and a Dirichlet condition for the PDE at $I_0$. 
We aim to apply an appropriate flux of particles to and from $\Omega_c$ based on the PDE profile across the interface $I_1$, which will ensure that the gradients of the different modelling regimes agree.
Feasibly, if this were the only condition, situations could arise where the gradients of the two regimes agree, but there is a notable discontinuity in the values of the density between descriptions.
To prevent this, we enforce a boundary condition on the PDE requiring the density on the lattice point at $I_0$ to match an average of the density of the surrounding compartments. 

First consider the Dirichlet matching condition at $I_0$, where we specify the PDE density in terms of particle numbers: 
\begin{equation} \label{I_0condition}
u_1(t+\Delta t) = (A_{-1}(t+\Delta t) + A_{1}(t+\Delta t))/2. 
\end{equation}
Writing this in terms of the analogous probability densities, we have:
\begin{equation} \label{dirichelet}
p_p(I_0,t+\Delta t) = \frac{p_c(I_0 + \frac{h}{2} , t+ \Delta t ) + p_c(I_0-\frac{h}{2} , t+ \Delta t )}{2} .
\end{equation}
Extending $p_c$ to continuous space and Taylor expanding the terms on the right-hand side (RHS) of equation \eqref{dirichelet} to first order, we find that:
\begin{align*}
p_p(I_0,t+\Delta t)  &= \Big[ p_c\left(I_0,t+\Delta t\right) + \frac{h}{2} \frac{\partial }{\partial x} p_c\left(I_0,t+\Delta t\right) + p_c\left(I_0,t+\Delta t\right) \\
 &- \frac{h}{2} \frac{\partial }{\partial x} p_c\left(I_0,t+\Delta t\right) + \mathcal{O}(h^2) \Big] /2 , \\
  &\approx  p_c\left(I_0,t+\Delta t\right),
\end{align*}
This suggests that matching condition \eqref{I_0condition} ensures agreement between the solution of the PDE and compartment-based particle numbers at $I_0$. The agreement will become exact in the limit $h\rightarrow 0$.

For the condition at $I_1$, we want to match the flux across $I_1$ in the compartment-based regime to that in the PDE regime. We will show that by enforcing the matched flux condition, the probability density for the compartment-based region evolves according to the diffusion equation in the limit of small compartment size.

We begin by writing down the mater equation \citep{gardiner1985handbook} for the probability density of a single particle at compartment $-m$, adjacent to the interface, $I_1$:
\begin{equation} \label{mastereqn}
p_{c}^{-m}(t+\Delta t) = \frac{D\Delta t}{h^2} \hspace{2pt} p_{c}^{-(m-1)}(t) + \left( 1-\frac{D\Delta t}{h^2} \right) p_{c}^{-m}(t) + \psi_p,
\end{equation}
where $p_{c}^{k}(t)$ is shorthand for $p_c(I_0+(2k+1)h/2,t)$ and describes the probability density for a single diffusing particle to be found in the $k$th compartment at time $t$. Here, $\psi_p$ is the flux imposed (as part of the hybrid algorithm) on compartment $-m$ from the right. 
If there were compartments to the right of the compartment labelled $-m$ (i.e. $-(m+1)$ etc (see Figure \ref{figDomain})) the true net flux would simply be
\begin{equation}
 \psi_c=\frac{D\Delta t}{h^2}(p_{c}^{-(m+1)}(t)-p_{c}^{-m}(t)).
\end{equation}
Instead we must approximate the true flux, $\psi_c$, by an ansatz derived from the PDE, $\psi_p$ as follows.

Suppose that the $l$th lattice point of the PDE lies on the interface $I_1$, and $w$ is the ratio of spacing between the compartment size, $h$, in $\Omega_c$ and the PDE finite difference lattice size, $\Delta x_p=(1-I_0)/K$, such that $w=\frac{h}{\Delta x_p}$.

We can interpolate the density in $\Omega_p$ at the centre of the $-m$th compartment by
\begin{equation*}
p^- = \left( 1 + \left\lfloor\frac{w}{2}\right\rfloor - \frac{w}{2}\right) p_p^{a_-}(t) + \left(\frac{w}{2} - \left\lfloor\frac{w}{2}\right\rfloor\right) p_p^{b_-}(t).
\end{equation*}  
where $ a_{-} = l - \left\lfloor \frac{w}{2}\right\rfloor,b_{-} = l - \left\lfloor\frac{w}{2}\right\rfloor - 1$.
Imagine an extra compartment $-(m+1)$ to the right of $I_1$. We could interpolate the density at the centre of this ghost compartment using a similar expression:
\begin{equation*}
p^+ = \left(1 + \left\lfloor\frac{w}{2}\right\rfloor - \frac{w}{2}\right) p_p^{a_+}(t) + \left(\frac{w}{2} - \left\lfloor\frac{w}{2}\right\rfloor\right) p_p^{b+}(t).
\end{equation*}  
where $ a_{+} = l + \floor{\frac{w}{2}},b_{+} = l + \floor{\frac{w}{2}} + 1.$

Given these interpolations of the PDE density at the centre of compartments we can approximate the diffusive flux across the interface and consequently set 
\begin{equation}
\psi_p=\frac{D\Delta t}{h^2} (p^+-p^-)\label{equation:psi_p}.
 \end{equation}

Substituting this into equation \eqref{mastereqn} gives
\begin{align} \label{mastereqn_with_psi}
p_{c}^{-m}(t+\Delta t) = p_{c}^{-m}(t) + \frac{D\Delta t}{h^2} \left( p_{c}^{-(m-1)}(t) -p_{c}^{-m}(t)+p^+-p^- \right).
\end{align}
Upon rearrangement this implies
\begin{align}
\frac{p_c \left(I_1- \frac{h}{2}, t+\Delta t \right) - p_c \left( I_1- \frac{h}{2}, t \right) }{\Delta t} =& \frac{D}{h^2} \left[ p_c \left(I_1 -\frac{3h}{2}, t \right) - p_c\left(I_1-\frac{h}{2},t\right)\right.\nonumber\\
 & - \left.p_p\left( I_1-\frac{h}{2}, t\right) + p_p\left(I_1+\frac{h}{2}, t\right) \right].\nonumber
\end{align}

In order to demonstrate the veracity of our choice of $\psi_p$, we extend $p_c$ to be a continuous function of space and Taylor expand terms on the RHS in space about the centre of the $-m$th compartment (i.e. $I_1-h/2$). 
Taylor expanding $p_c \left(I_1- \frac{h}{2}, t+\Delta t \right)$ in time and taking the diffusive limit, we find we recapitulate the diffusion equation for the probability density at $I_1-h/2$ if $p_p(I_1-h/2), t + p_p(I_1+h/2) = \allowbreak p_c(I_1-h/2, t) + p_c(I_1+h/2)$ or equivalently $\psi_p=\psi_c$. 
Consequently, this indicates that the flux $\psi_p$ given by equation \eqref{equation:psi_p} is an appropriate boundary condition for the compartment-based model.

Given our two matching conditions at either end of the interface, the hybrid algorithm can be implemented in a time-driven sense as follows:

\noindent 
\begin{minipage}{\textwidth} 
  \centering 
\begin{center}
 \begin{tabular}{|p{12cm}|}  
\hline
  (i) Initialise number of particles $A_k, \hspace{5pt}$ $k=-m,...,-1,1,...,n$ in compartments in $\Omega_c $ and apply consistent initial conditions in $\Omega_p $. \\
  (ii) Select the compartment-based time step $\Delta t$, such that the probability of more than one event occurring per time-step is $O(\Delta t^2)$, and a maximum duration of the simulation, $T_{final}$. Set $t:=0$.\\
 (iii) Calculate $\psi = N \psi_p$, where $\psi_p$ is as in equation \eqref{equation:psi_p}. \\
  Draw a uniform random number $r_1$. \\
  If $r_1 < | \psi | $, then update $A_{-m}(t):=A_{-m}(t)+\sgn{(\psi)} $. \\
  (iv) Calculate a uniform random number $r_2$.\\
  If $r_2 < \alpha_0 \Delta t$  where $\alpha_0 $ is the total propensity of the `reaction' events in the compartment-based regime, then a reaction occurs in that time step. \\
 (v) If a `reaction' occurs, generate a uniform random number $r_3 $, and find $j$ such that 
   $\sum\limits_{i=1}^{j-1} \alpha_i \le r_3 \alpha_0 < \sum\limits_{i=1}^{j} \alpha_i$.\\
Update number of particles in each compartment according to chosen reaction, $j$. \\
(vi) Update time such that $t:=t+\Delta t$.\\
(vii) Update PDE region $\Omega_p $ using an appropriate numerical method.
Apply the boundary condition at the right-hand boundary and the coupling condition at $I_0$ as follows: \\
 $u_1(t+\Delta t) = (A_{-1}(t+\Delta t) + A_{1}(t+\Delta t))/2. $\\
(viii) If $t<T_{final} $, then go back to step (iii). Else end. \\
\hline 
\end{tabular}

\end{center}

  \medskip 
 Algorithm 1: Time-based hybrid algorithm for stochastic reaction-diffusion simulations using a compartment-based region and an overlapping PDE-based region. 
\end{minipage}

Note the factor of $N$ in the calculation of $\psi$ at step (iii) is due to the scaling between concentration and the probability distribution for a single particle. 
Both ``time-based'' and ``event-based'' versions of the hybrid coupling algorithm are possible.
The main difference between these is that the time-based algorithm uses a fixed time step $\Delta t$ to update both $\Omega_c$ and $\Omega_p$, while the event-based algorithm steps forward to the next reaction in $\Omega_c$, while still fixing a maximum time step in $\Omega_p$ for updating the PDE.
For systems with large numbers of particles, the event-based algorithm will be more efficient as it allows the use of larger time steps in the stochastic regime so fewer steps of the algorithm are required.
However, for simplicity, we present here the time-based version, Algorithm 1. 

\section{Results}

\subsection{Numerical simulations}\label{section:numerical_simulations}
\subsubsection{Test problem: diffusion}
We will begin our examination of practical applications of the hybrid coupling algorithm by applying the method to a test problem in which particles diffuse with diffusion constant $D$.
With large copy numbers of particles in the system, the density of diffusing particles, $u(x,t)$, is governed by the diffusion equation:
\begin{equation}
  \frac{\partial u}{\partial t} = D \frac{\partial ^2 u}{\partial x^2}, \hspace{12pt} x \in \Omega\label{equation:diffusion_equation} .
\end{equation}
Adding reactions to this system should not affect the boundary behaviour directly and therefore it is sufficient to test our model on a problem of this type \citep{erban2007reactive}.
As previously specified (but without loss of generality), our domain is $\Omega = [-1,1] $ with zero flux boundary conditions at both ends. 
This domain is divided into a deterministic PDE-based region and a stochastic compartment-based region as required by the hybrid coupling algorithm.
We choose $\Omega_c = [-1,0.1], \Omega_p=[0,1]$.
The left hand interface of the overlap region is at $I_0 = 0$ while the right hand interface of the overlap region lies at $I_1 = 0.1$.

We consider three different initial conditions, $\phi(x)$: a uniform initial condition, demonstrating that the algorithm can maintain an equilibrium state, a step function with all the mass in $[0,1]$, that is: 
\begin{equation} \label{RHS_IC}
\phi(x)=N . \mathds{1} _{x \ge 0} = \begin{cases} 0, & x < 0, \\ N, & x \ge 0, \end{cases}  
\end{equation} 
 and a step function with all the mass in $[-1,0]$, that is: 
\begin{equation} \label{LHS_IC}
\phi(x)=N . \mathds{1} _{x \le 0} = \begin{cases} N, & x \le 0, \\ 0, & x > 0. \end{cases}  
\end{equation}
These provide a robust test of our hybrid algorithm in a variety of different scenarios, showing it can maintain net flux from each region to the other. 

We have performed simulations of the hybrid model, using the three different initial conditions described above. 
We also present the analytical solutions of the mean-field diffusion equation.
In particular, suppose that all the mass is initially in $[0,1]$, as in \eqref{RHS_IC}.
Using a Green's function and an infinite series of images at the boundaries we obtain an analytical solution to equation \eqref{equation:diffusion_equation} of the form:
\begin{align}
u(x,t) &= \frac{N}{2} \erf{\left( \frac{1 -x}{\sqrt{4 D t}} \right) } - \frac{N}{2} \erf{\left( \frac{-x}{\sqrt{4 D t}}\right) } \nonumber \\ 
&+ \frac{N}{2} \sum_{k=1}^{\infty} \bigg\{  \erf{\left( \frac{1+(x-2k)(-1)^{k+1}}{\sqrt{4 D t}} \right) } - \erf{\left( \frac{(x-2k)(-1)^{k+1}}{\sqrt{4 D t}} \right)} \nonumber \\
&+ \erf{\left( \frac{1+(x+2k)(-1)^{k+1}}{\sqrt{4 D t}}\right) } - \erf{\left( \frac{(x+2k)(-1)^{k+1}}{\sqrt{4 D t}}\right) } \bigg\}  ,\label{erfsoln}
\end{align}
where we have written the solution in terms of error functions.
The solution for the initial condition of a step function with all the mass in $[-1,0]$ (as in equation \eqref{LHS_IC}) can be obtained by symmetry from equation (\ref{erfsoln}). This solution is used in both Figure \ref{figure:LHS_gradient} and Figure \ref{figure:RHS_gradient}. 

\begin{figure}[h!]
\begin{center}
\subfigure[]{
\includegraphics[width=0.31\columnwidth]{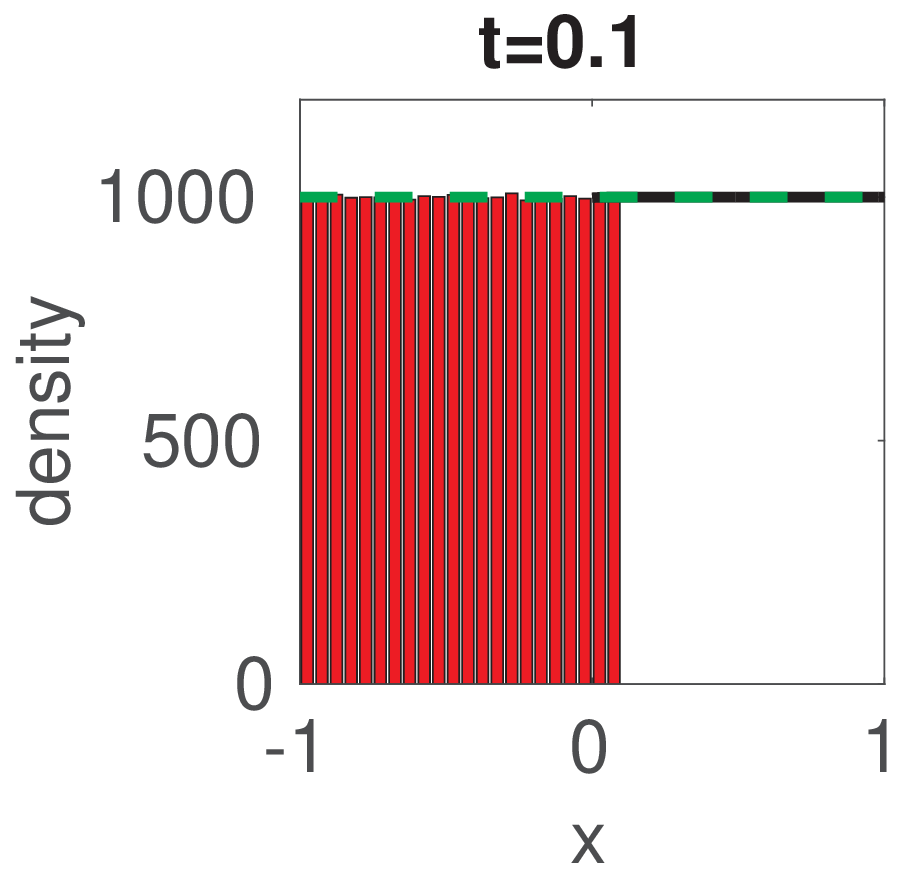}
\label{figure:uniform_initial_condition}
}
\subfigure[]{
\includegraphics[width=0.31\columnwidth]{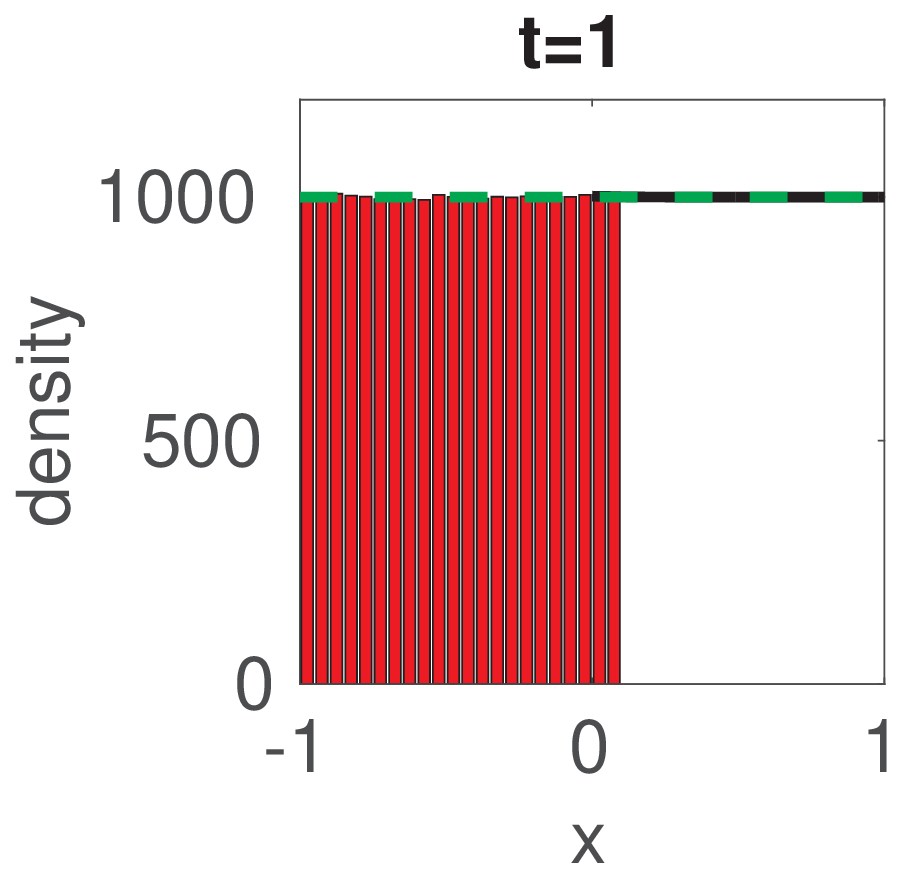}
\label{figure:uniform_60_algorithm_1}
}
\subfigure[]{
\includegraphics[width=0.31\columnwidth]{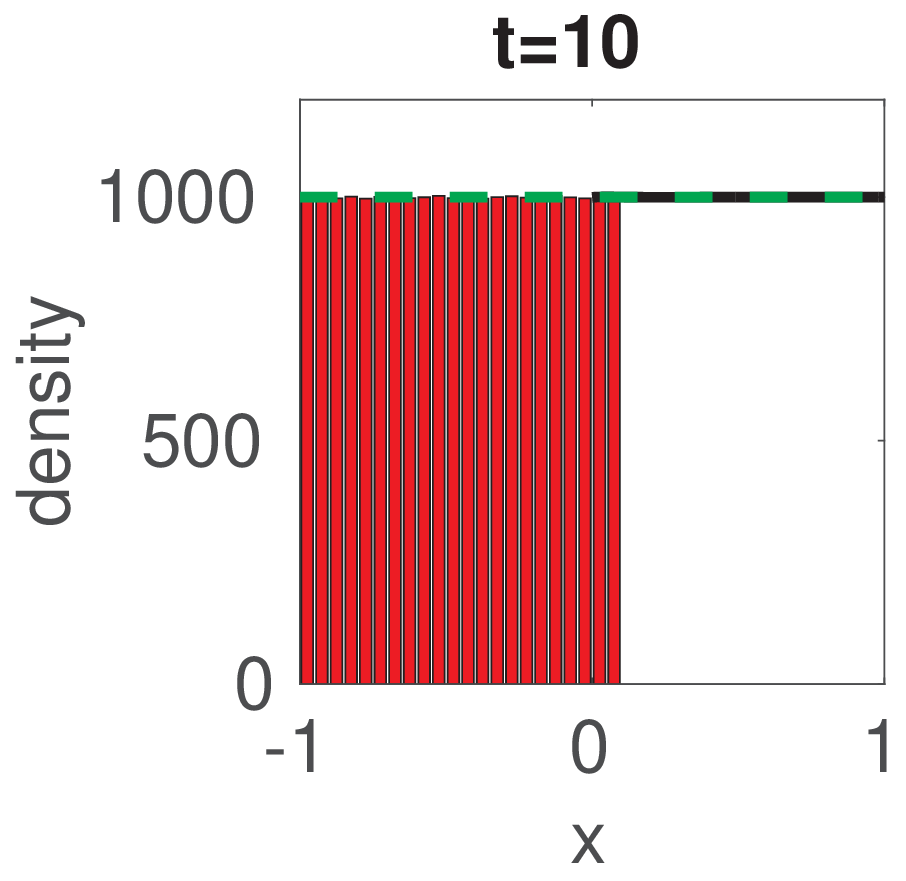}
\label{figure:uniform_60_algorithm_2}
}
\end{center}
\caption[Diffusion with mass initially uniform]{Simulating simple diffusion starting from a uniform distribution of mass throughout the domain $\Omega$. Panels \subref{figure:uniform_initial_condition},
\subref{figure:uniform_60_algorithm_1} and \subref{figure:uniform_60_algorithm_2} show the particle density at times $t=0.1$, $t=1$ and $t=10$ respectively. Simulations are performed using the hybrid coupling algorithm set out in Algorithm 1. Parameters used are $D=0.025, \Delta t=0.001, h=0.05, \Delta x_p=0.01$ and the simulation results are averaged over 100 repeats. The black line represents the density in $\Omega_p $ and the red bars represent the particle density in $\Omega_c$.
The dashed green line shows the (trivial) analytic solution. 
}
\label{figure:uniform_gradient}
\end{figure}

\begin{figure}[h!]
\begin{center}
\subfigure[]{
\includegraphics[width=0.31\columnwidth]{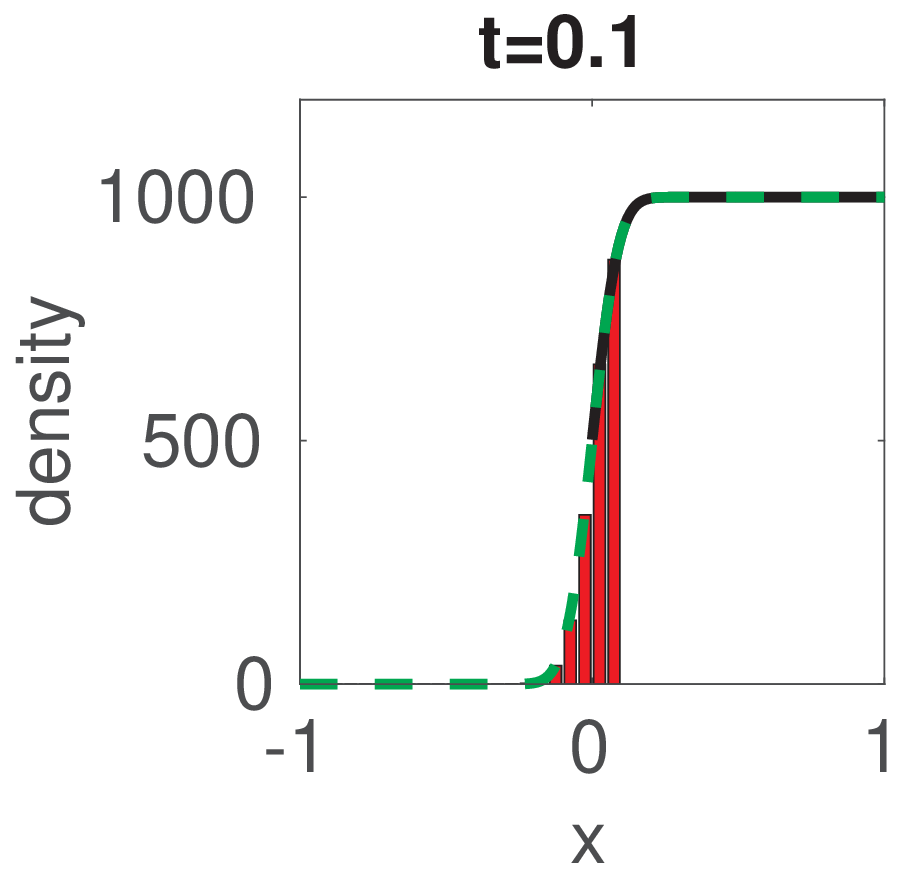}
\label{figure:LHS_initial_condition}
}
\subfigure[]{
\includegraphics[width=0.31\columnwidth]{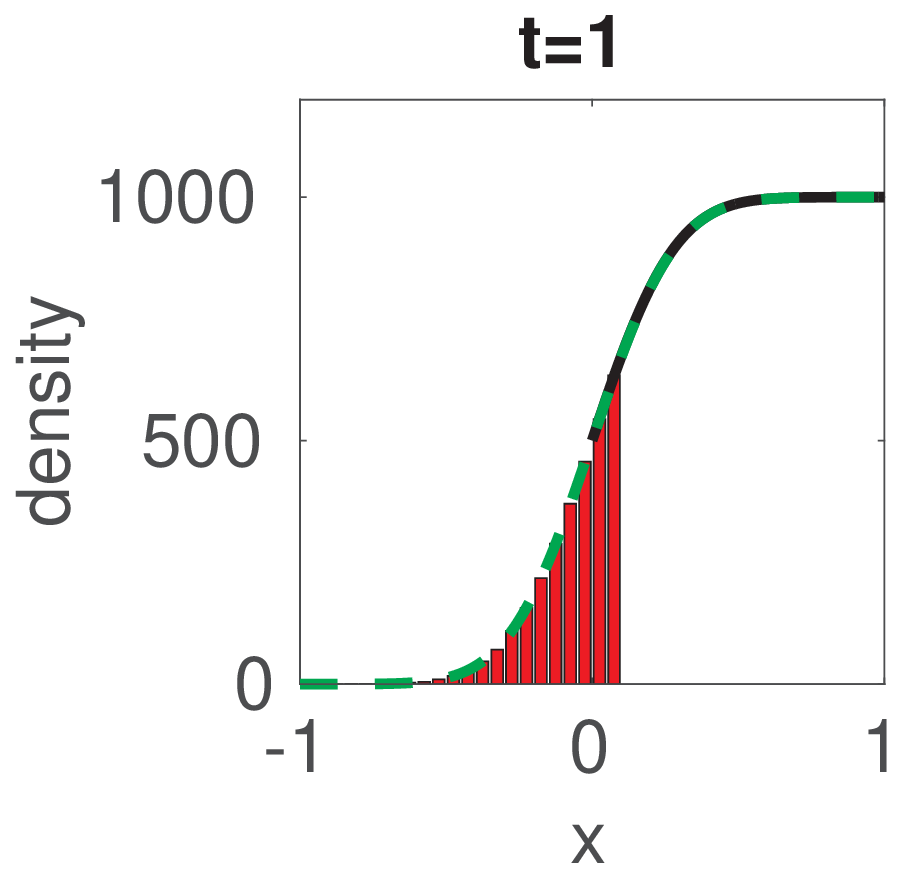}
\label{figure:LHS_60_algorithm_1}
}
\subfigure[]{
\includegraphics[width=0.31\columnwidth]{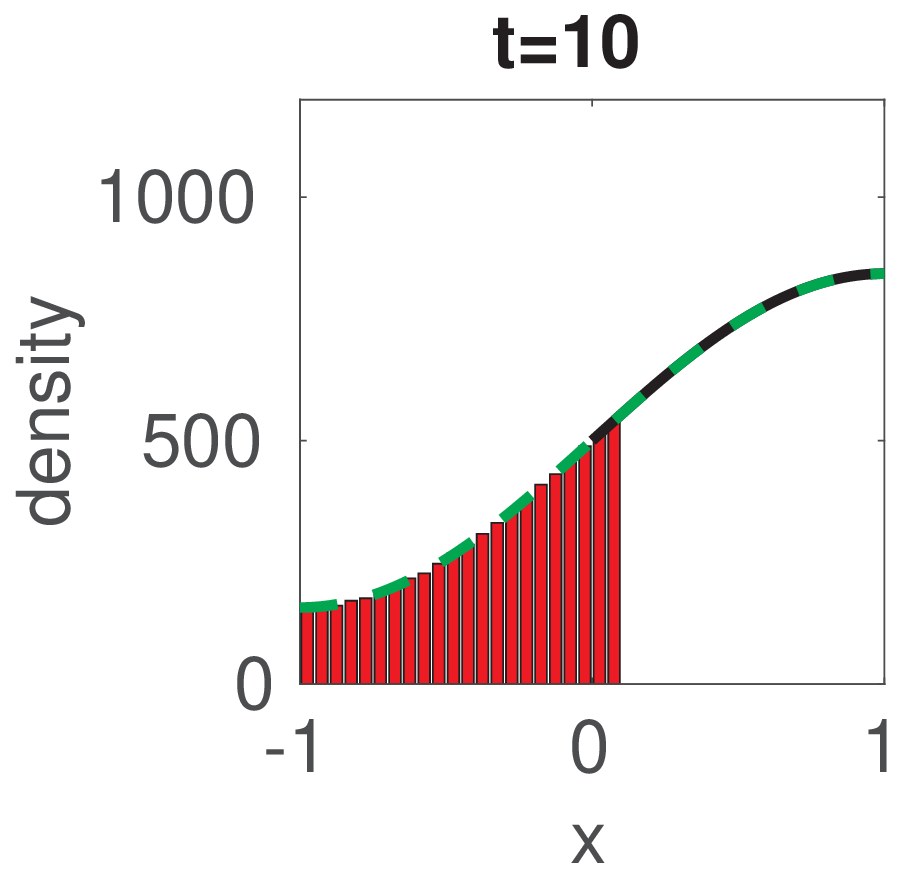}
\label{figure:LHS_60_algorithm_2}
}
\end{center}
\caption[Diffusion with mass initially in the stochastic region]{Simulating simple diffusion starting from a step function with mass in $[0,1]$. Panels \subref{figure:LHS_initial_condition},
\subref{figure:LHS_60_algorithm_1} and
\subref{figure:LHS_60_algorithm_2} show the particle density at times $t=0.1$, $t=1$ and $t=10$ respectively. Simulations are performed using the hybrid coupling algorithm set out in Algorithm 1. Parameters, repeats and figure descriptions are as for Figure \ref{figure:uniform_gradient}.}
\label{figure:LHS_gradient}
\end{figure}

\begin{figure}[h!]
\begin{center}
\subfigure[]{
\includegraphics[width=0.31\columnwidth]{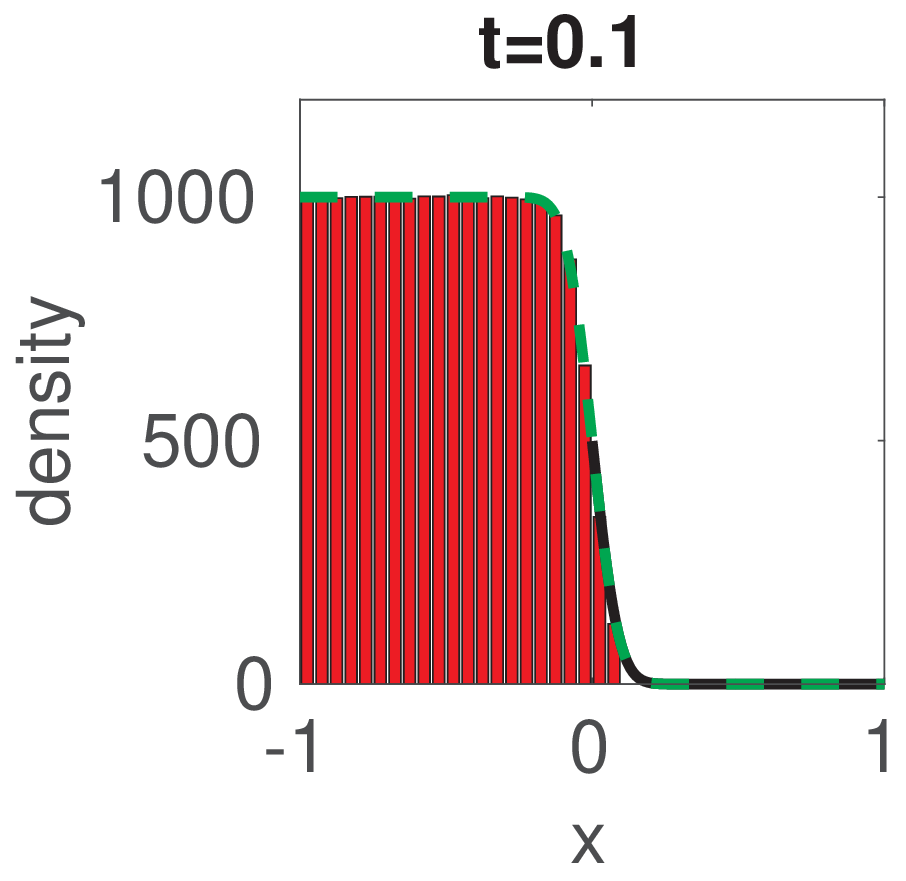}
\label{figure:RHS_initial_condition}
}
\subfigure[]{
\includegraphics[width=0.31\columnwidth]{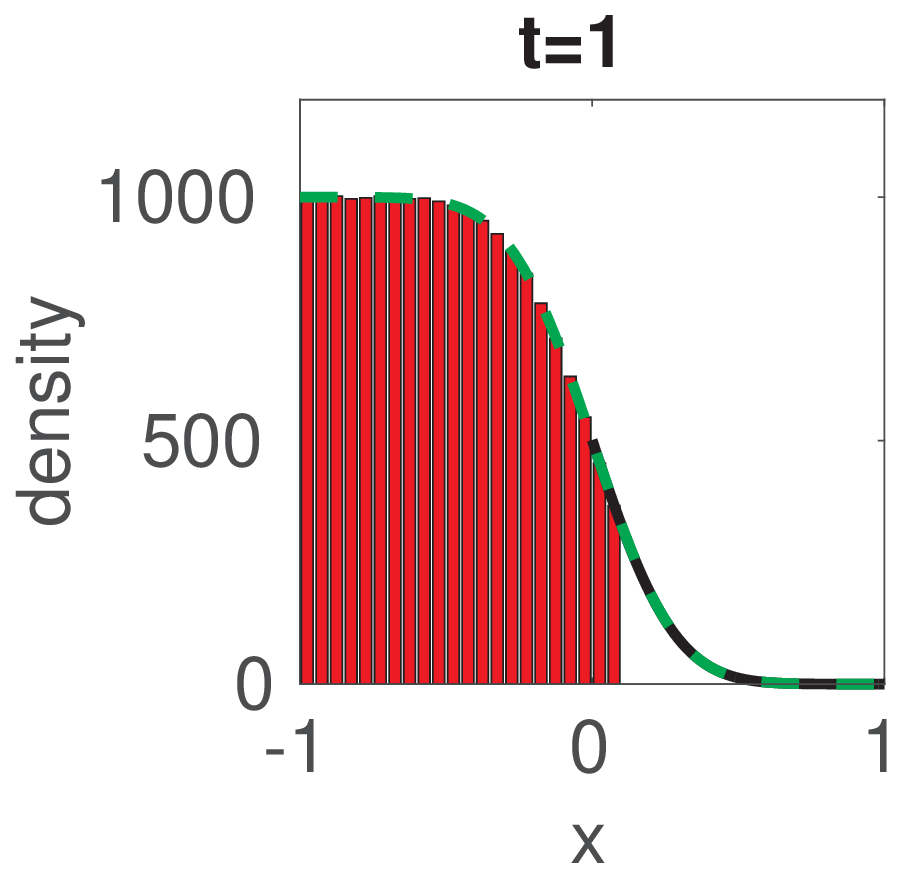}
\label{figure:RHS_60_algorithm_1}
}
\subfigure[]{
\includegraphics[width=0.31\columnwidth]{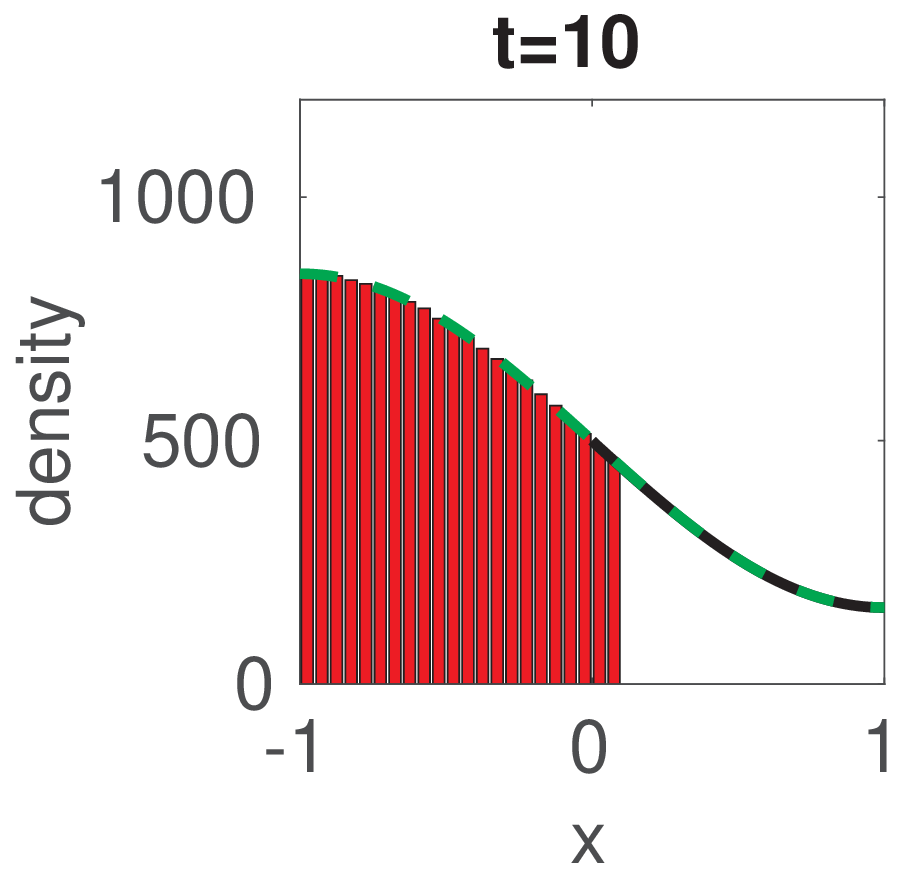}
\label{figure:RHS_60_algorithm_2}
}
\end{center}
\caption[Diffusion with mass initially in the deterministic region]{Simulating simple diffusion starting from a step function with mass in $[-1,0]$. Panels \subref{figure:RHS_initial_condition},
\subref{figure:RHS_60_algorithm_1} and
\subref{figure:RHS_60_algorithm_2} show the particle density at times $t=0.1$, $t=1$ and $t=10$ respectively. Simulations are performed using the hybrid coupling algorithm set out in Algorithm 1. Parameters, repeats and figure descriptions are as for Figure \ref{figure:uniform_gradient}.}
\label{figure:RHS_gradient}
\end{figure}

Comparisons between our hybrid model and the mean-field analytical solution are shown in Figures \ref{figure:uniform_gradient}, \ref{figure:LHS_gradient}, and \ref{figure:RHS_gradient} for a range of times. Agreement is observed between the simulated results and the analytic solutions. 

\begin{figure}[h!]
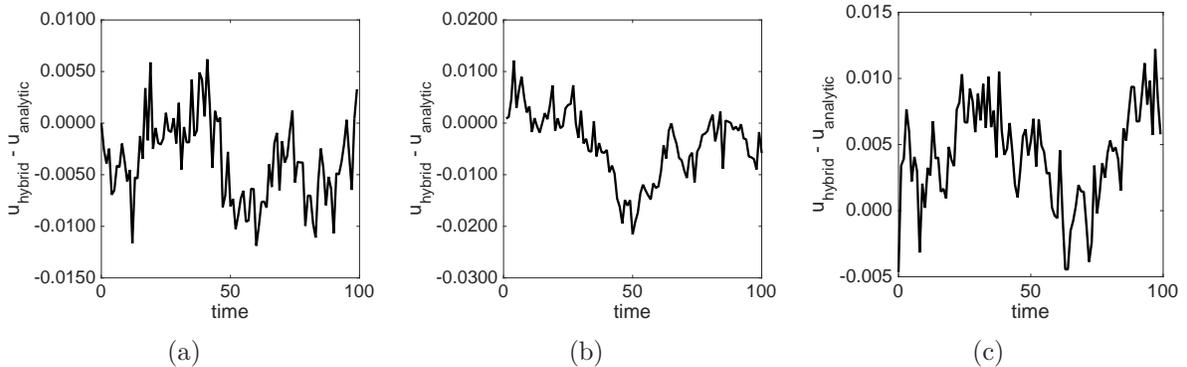

\begin{center}
\subfigure[]{
\includegraphics[width=0.3\columnwidth]{error_over_time_IC3.eps}
\label{figure:error1}
}
\subfigure[]{
\includegraphics[width=0.3\columnwidth]{error_over_time_IC2.eps}
\label{figure:error2}
}
\subfigure[]{
\includegraphics[width=0.3\columnwidth]{error_over_time_IC1.eps}
\label{figure:error3}
}
\end{center}
\caption[Evolution of error]{The evolution over time of the error obtained from simulations using the hybrid method with parameters as in Figure \ref{figure:RHS_gradient}. Panel
\subref{figure:error1} employed a uniform distribution of mass throughout the domain as the initial condition, panel
\subref{figure:error2} a step function with mass in $\Omega_p$, and panel \subref{figure:error3} a step function with mass in $\Omega_c$.
The error is calculated as the difference between the average density given by the hybrid model over 100 repeats and the deterministic expected value of the density.}
\label{errorplot}
\end{figure}

Quantitative comparisons of the simulations from the hybrid model with the analytic solutions can be seen in Figure \ref{errorplot}.
We compute the error as a sum across the entire spatial domain $\Omega$ of absolute values of the difference between the average of the hybrid model and the analytic mean field solutions. 
This difference is computed at the centre of each region of width $h$, in both $\Omega_c$ and $\Omega_p$. 
The resulting stochastic error is normalised by the total number of particles in the system.
The errors are unbiased about 0, and crucially, in each case, the magnitude of the absolute error does not increase over time.
This demonstrates quantitatively the agreement between the two modelling regimes.

\subsubsection{Test problem: morphogen gradient}
We also apply our model to another test problem: the formation of a morphogen gradient.
For this problem, we use the same domain and partitioning as before. 
Morphogen molecules are produced at rate $J$ at $x=1$ and throughout the domain morphogen molecules decay with constant rate $\mu $ and diffuse with diffusion coefficient $D$.
When there are sufficiently many molecules in the system, we expect the density of molecules, $u(x,t)$, to be governed by the following PDE:
\begin{equation} \label{morpheqn}
  \frac{\partial u}{\partial t} = D \frac{\partial ^2 u}{\partial x^2} -\mu u + J \delta (x-1), \hspace{12pt} x \in \Omega.
\end{equation}
We apply zero flux conditions at the boundaries and initially we assume there are no molecules in the system.

\begin{figure}[h!]
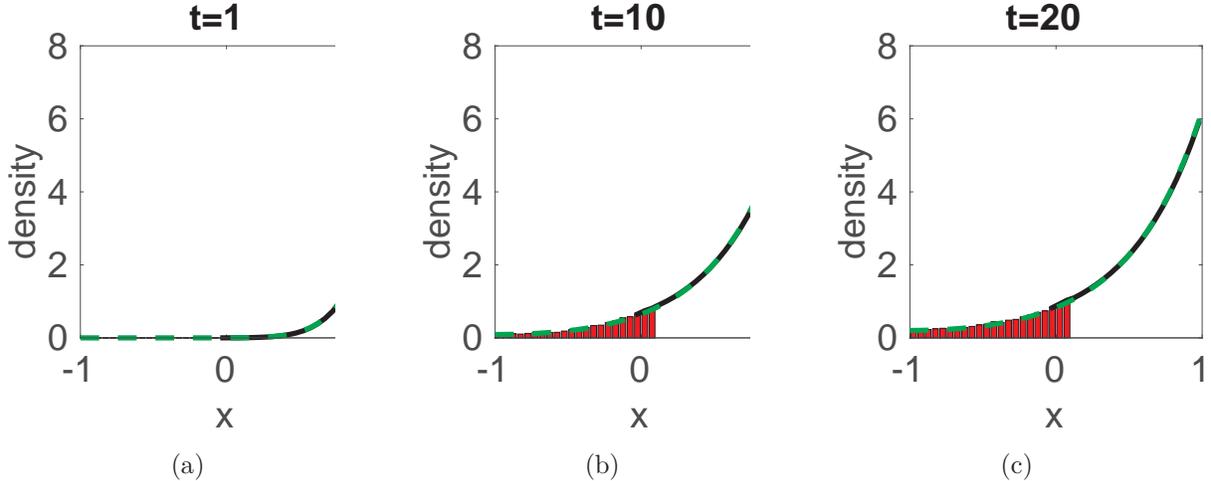

\begin{center}
\subfigure[]{
\includegraphics[width=0.31\columnwidth]{morphogen_gradient_t1.eps}
\label{figure:morphogen_initial_condition}
}
\subfigure[]{
\includegraphics[width=0.31\columnwidth]{morphogen_gradient_t10.eps}
\label{figure:morphogen_60_algorithm_1}
}
\subfigure[]{
\includegraphics[width=0.31\columnwidth]{morphogen_gradient_t20.eps}
\label{figure:morphogen_60_algorithm_2}
}
\end{center}
\caption[The evolution of a morphogen gradient over time]{Averaged simulations of the morphogen gradient system with the hybrid algorithm (using equation \eqref{morpheqn} in $\Omega_p$ and stochastic simulations of the reaction scheme in $\Omega_c$) compared to the analytical solution of \eqref{morpheqn}. Initially the domain is empty. 
Simulations are performed up to $t=20$ and averaged over 100 repeats with parameters $D=0.05,  \mu=0.2, J=125, h=0.05, \Delta x_p=0.01$. 
Panels \subref{figure:morphogen_initial_condition}, \subref{figure:morphogen_60_algorithm_1} and \subref{figure:morphogen_60_algorithm_2} show the particle density at $t=1$, $t=10$ and $t=20$ respectively. Figure descriptions are as for Figure \ref{figure:uniform_gradient}.}
\label{figure:morphogen_gradient}
\end{figure}

The results of simulating this morphogen system are shown in Figure \ref{figure:morphogen_gradient}.
The system was simulated up until $t=20$ after which point the system had approached steady state.
Good agreement can be seen between the hybrid simulation algorithm and the analytical solution of \eqref{morpheqn}.

\subsubsection{Test problem: travelling wave} \label{section:travelling}
The occurrence of travelling waves is common throughout the natural world: they describe a variety of phenomena from propagation of genes in a population \citep{fisher1937wave}, to epidemic outbreaks \citep{grenfell2001travelling}, and in the FitzHugh-Nagumo equations for a nerve axon pulse \citep{keener1998mathematical}.  

One commonly used model for a travelling wave front is the Fisher-KPP equation:
\begin{equation} \label{tweqn}
\frac{\partial u}{\partial t} = D \frac{\partial ^2 u}{\partial x^2} + k_1 u - k_2 u^2,
\end{equation}
where $D$ is the diffusion coefficient, and $k_1$ and $k_2$ are reaction rates.
This is a non-linear reaction-diffusion equation for the concentration or population density $u$ in one dimension. 
It can be shown that this results in the formation of a travelling front with a minimum wave-speed of $c=2\sqrt{D k_1}$, given continuous initial conditions with compact support \citep{murray2002mathematical}. 

Consider the reversible chemical reaction: 
\begin{eqnarray} \label{twreaction}
\ce{A 
<=>[k_1][k_2]
{ A}  + { A } }.
\end{eqnarray}
Using the law of mass action in a deterministic setting \citep{murray2002mathematical} and including diffusion effects, results in the Fisher-KPP equation \eqref{tweqn} as a description of the evolution of the chemical concentration.
To investigate stochastic simulations of the propagation of travelling waves, we can interpret the reaction system \eqref{twreaction} in a stochastic sense\citep{moro2004hybrid}.
The stochastic simulations of wave front propagation do not generally match the deterministic models, with stochastic models resulting in a different wave speed than given by the deterministic model and different speeds given depending on the stochastic model used \citep{robinson2013adaptive}. The wave speed in the stochastic models approaches that of the deterministic model in the continuum limit of many particles, but does so relatively slowly with  $v=v_{min} - \frac{K}{(\ln{N})^2 }$, where $v_{min}$ is the minimum velocity for the deterministic model, $K$ is a constant, and $N$ is the total number of molecules \citep{brunet1999microscopic}.
By considering moments of the appropriate chemical master equation, we obtain a hierarchy of coupled equations, where the $k\rm{th}$ moment depends upon the $(k+1)\rm{th}$ moment \citep{breuer1994fluctuation}. In order to obtain a closed system we must make a closure approximation.
The degree of agreement between the deterministic and stochastic descriptions will depend on the validity of this closure assumption.

We note that the nature of reaction scheme \eqref{twreaction} means that population growth in compartments ahead of the wave front does not begin until there is at least one particle present in that compartment.
The discretisation of particles in the stochastic model, therefore, restricts the progress of the wave and results in the lower wave speed in comparison to the deterministic interpretation \citep{panja2004effects}.

Given that we do not expect the stochastic model to correspond to the deterministic model in the mean-field
we will use a fully stochastic compartment-based description of the system for comparison with our hybrid system in order to determine its accuracy (as opposed to the PDE description which represented the mean-field behaviour of the previous test systems). 
We expect to make computational savings by using a PDE to describe the mean field behaviour behind the wave whist using the stochastic compartment-based model to simulate behaviour at the wave front and ahead of the wave, which determines the wave speed.

Applications of hybrid models to travelling waves have been made in previous work.
\citet{moro2004hybrid} have successfully demonstrated such a model, using a flux based approach similar to that of \citet{flekkoy2001coupling}.
This hybrid model was then used to confirm the scaling of the velocity correction for the stochastic mesoscopic model.
Further to this, an adaptive version of the two-regime method has also been applied to a travelling wave problem \citep{robinson2013adaptive}.
This model couples a microscopic Brownian motion based description to a mesoscopic compartment-based description, as in the original two regime method \citep{flegg2012two}.
In addition, the interface between the two regions is, in this case, allowed to move adaptively following the propagation of the front \citep{robinson2013adaptive}. 
This enables the microscopic description to represent the most appropriate region of the domain, following the front of the wave, with the less computationally intensive mesoscopic description remaining behind the wave. 

\begin{figure}[h!]
\begin{center}
\subfigure[]{
\includegraphics[width=0.31\columnwidth]{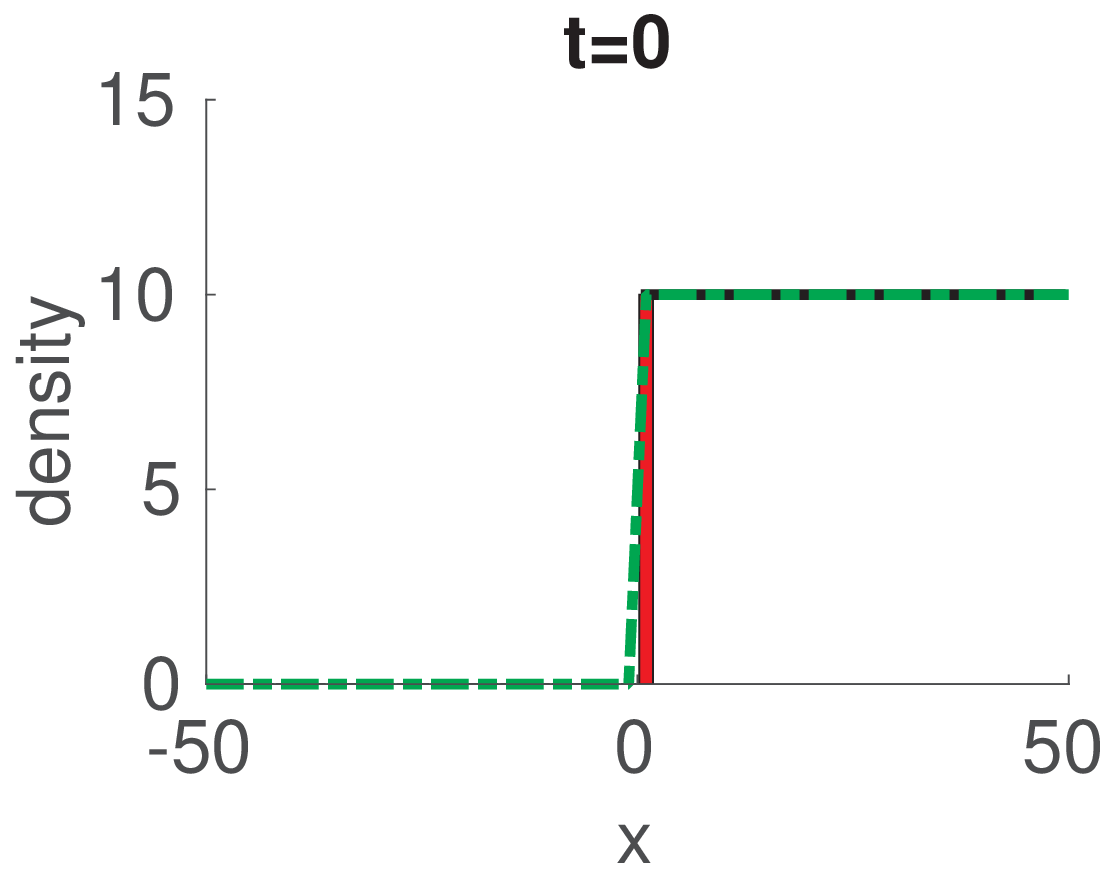}
\label{figure:tw_initial_condition}
}
\subfigure[]{
\includegraphics[width=0.31\columnwidth]{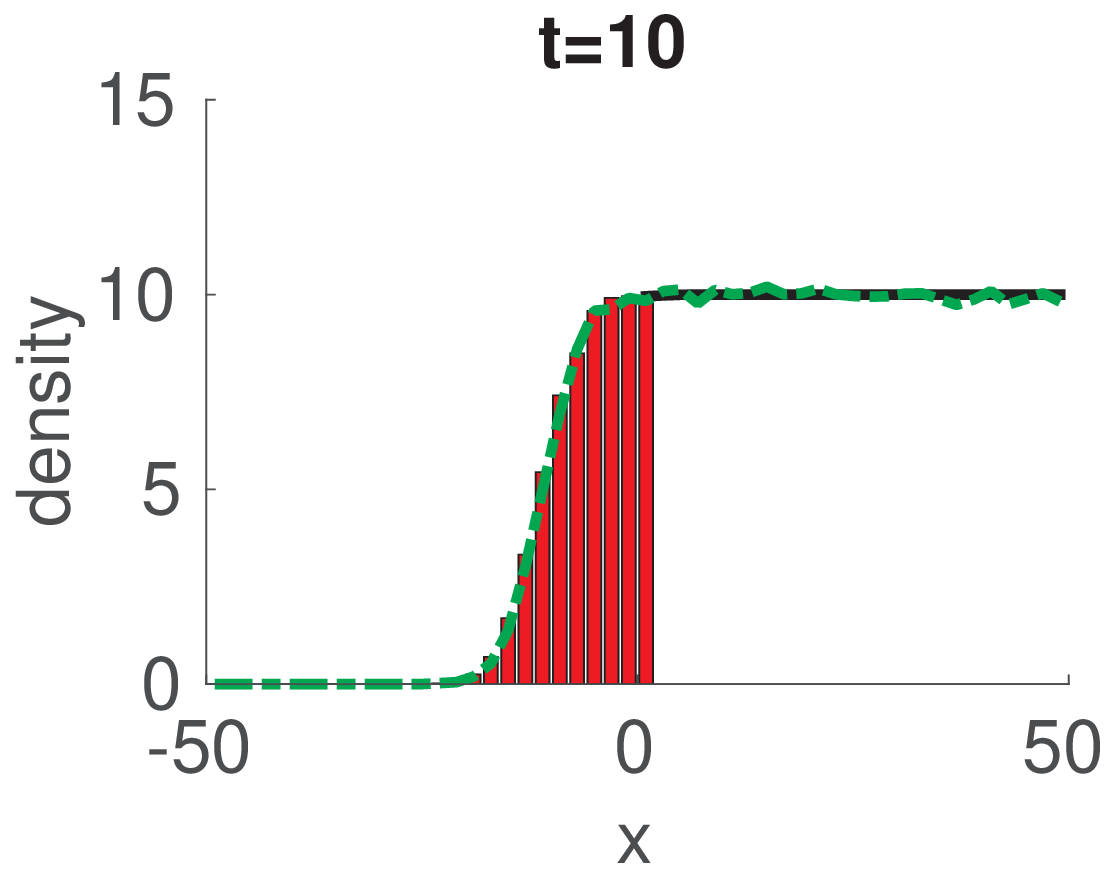}
\label{figure:tw_60_algorithm_1}
}
\subfigure[]{
\includegraphics[width=0.31\columnwidth]{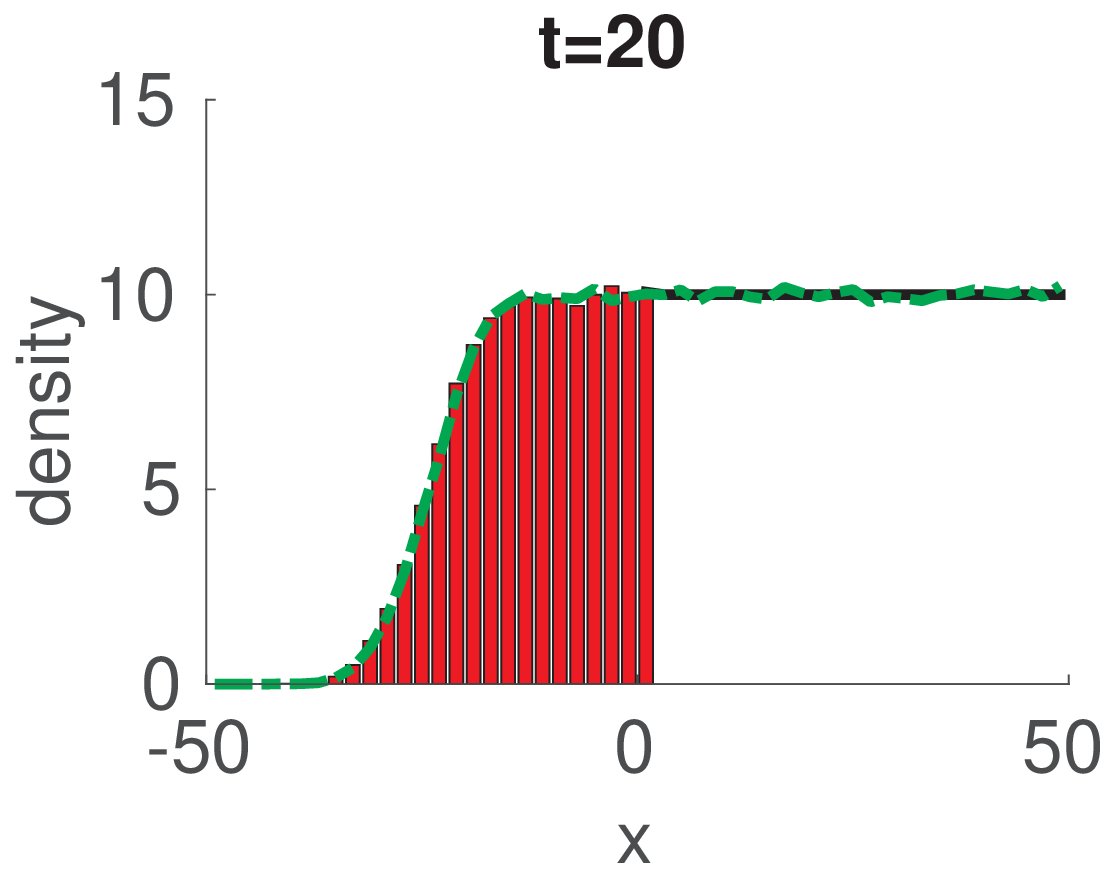}
\label{figure:tw_60_algorithm_2}
}
\end{center}
\caption{Simulating a travelling wave using the hybrid model, and the fully stochastic scheme \eqref{twreaction}.
The results shown have been averaged over 1000 repeats. Parameters used are $D=1, h=2, \Delta x_p=0.5, k_1=1, k_2=0.1 $.
Panels \subref{figure:tw_initial_condition}, \subref{figure:tw_60_algorithm_1} and
\subref{figure:tw_60_algorithm_2} show the particle density at times $t=0$, $t=10$ and $t=20$ respectively. The green dashed line shows the result of fully stochastic simulations while the red histogram and black line shows the result of the hybrid model in the particle-based and PDE-based regions respectively.}
\label{figure:tw}
\end{figure}

We demonstrate that our hybrid model can be applied successfully to a travelling wave using a fixed overlap region between the models, taking the domain as $\Omega =[-L,L]$ where $L=50$, with an overlap region at $[0,2]$. 
Consequently we have $\Omega_c=[-50,2]$ while $\Omega_p=[0,50]$.
We take our initial condition as a step function: $\phi (x) =  10\cdot \mathds{1} _{x>0}$. 
The results of simulations are displayed in Figure \ref{figure:tw}, showing the close agreement between the hybrid model and the fully stochastic model. 
The hybrid model accurately captures the stochastic behaviour at the front of the wave that is missed by the fully PDE-based model.

An important measure when investigating stochastic simulations of reaction system \eqref{twreaction} is the resulting wave speed.
It can be difficult with a stochastic model to specify exactly where the wave front is at a given time and to quantify exactly how fast it is moving, since there will inevitable be noise in the results of simulations\citep{breuer1994fluctuation}.
We choose to use the method outlined by \citet{robinson2013adaptive}, which considers the rate of change of the total mass, $M(t)$ in the system.
For times $t_2$ and $t_1$, we take:
\begin{equation} \label{cest}
 \hat{c} = \frac{M(t_2)-M(t_1)}{t_2-t_1} \frac{k_2}{k_1},
\end{equation}
where the factor $\frac{k_2}{k_1}$ is necessary since the height of the wave will approach $\frac{k_1}{k_2}$. 
Dividing the rate of change of mass by this factor gives a measure of how fast the wave is propagating through the domain. 

\begin{figure}[h!]
\centering
\includegraphics[height=3in]{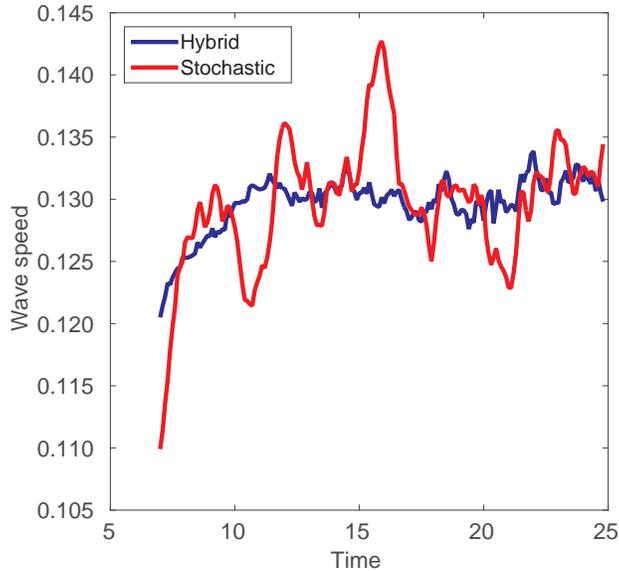}
\caption{A comparison of the wave-speeds resulting from simulations of a travelling wave using a fully stochastic model (shown in red/light grey) and the hybrid model (shown in blue/dark grey). 
The parameters for the simulations are the same as in Figure \ref{figure:tw}. 
The wave-speed was estimated using equation \ref{cest} at regular time intervals and smoothed with a moving average across five time units. }
\label{wavespeed}
\end{figure}

Figure \ref{wavespeed} shows a comparison between the wave speeds obtained from the fully stochastic compartment-based model, the hybrid model and the deterministic PDE model. There is more variation in the fully stochastic model since the PDE part of the hybrid model acts to dampen the fluctuations in the stochastic part of the model.
Good agreement is seen between the wave-speeds of the two models as estimated by a moving average of the wave-speed estimates, after an initial transient. 
The slower initial wave-speed observed in both models is explained by the steep initial condition, which first needs to approach the profile of the travelling wave before it starts to move at constant speed.

\subsubsection{Adaptive interface via a local detection criterion} \label{section:adaptive}

In certain situations, as with the travelling wave presented in the previous section, the region of interest with lower particle numbers changes position dynamically.
In order to capture most effectively the detail in this area whilst reducing the computational requirements it will be useful to have an interface that also changes position, so that regions with higher particle numbers can more often be modelled using the PDE. 
To ensure that the interface moves correctly, we initiate moves of the interface adaptively based on a detection condition of the particle density near the interface.
Moving the interface should also prevent unnecessary simulation of large particle numbers using the stochastic regime in regions where this is not required; for example, behind the wave front for larger times in the travelling wave model (see Figure \ref{figure:tw} panel 
\subref{figure:morphogen_60_algorithm_2}).

Such adaptive interfaces have previously been implemented in hybrid models \citep{spill2015hybrid} and in several works \citep{robinson2013adaptive, Ho2013multiscale} based on the previously mentioned two regime method \citep{flegg2012two}.
The two regime method implements a coupling between a compartment-based stochastic model and a molecular based stochastic model. 
In the adaptive two regime method \citep{robinson2013adaptive}, the interface between the two models moves adaptively in increments of the compartment width $h$. 
The moves are made to keep the density of particles below a certain threshold $u_{max}$. 
If the density of the particles in the compartment adjacent to the interface is above $u_{max}$, then the interface is moved into the compartment-based region. 
Conversely, if the density in the molecular region is below another threshold then the interface is moved into the molecular region. 
This threshold is chosen as $u_{max} - \delta u$, where $\delta u$ is a small (constant) increment, to prevent unnecessary fluctuations in the position of the interface due to the stochastic of the system \citep{Ho2013multiscale}. 
For similar reasons, the condition for updating the position of the interface is not checked every time step but after a fixed number of time steps to prevent errors resulting from moving the interface too frequently \citep{Ho2013multiscale}.

We choose to move the interface only by small increments equal to the compartment width $h$ after each successful check of a local detection criterion. 
This criterion is checked at intervals of $\eta $ steps of the algorithm. 
The requirement for moving the interface is that the density in both the compartment-based region and PDE-based region near the interface must be either above $u_{max}$ or below $u_{max}-\delta u$. 
Specifically, we check the compartments either side of interface $I_0$ in $\Omega_c$ and PDE points at equivalent positions either side of the interface $I_1$. 

\begin{figure}[h!]
\begin{center}
\subfigure[]{
\includegraphics[width=0.31\columnwidth]{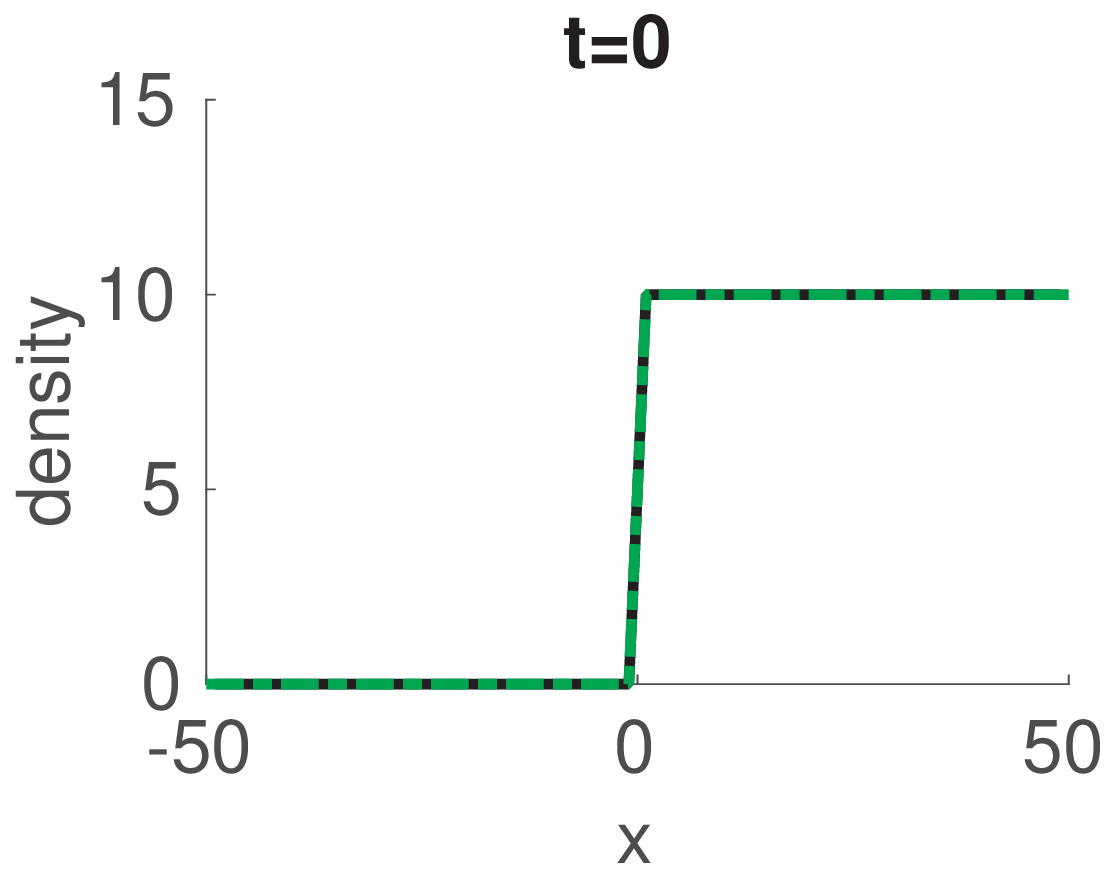}
\label{figure:tw_moving_initial_condition}
}
\subfigure[]{
\includegraphics[width=0.31\columnwidth]{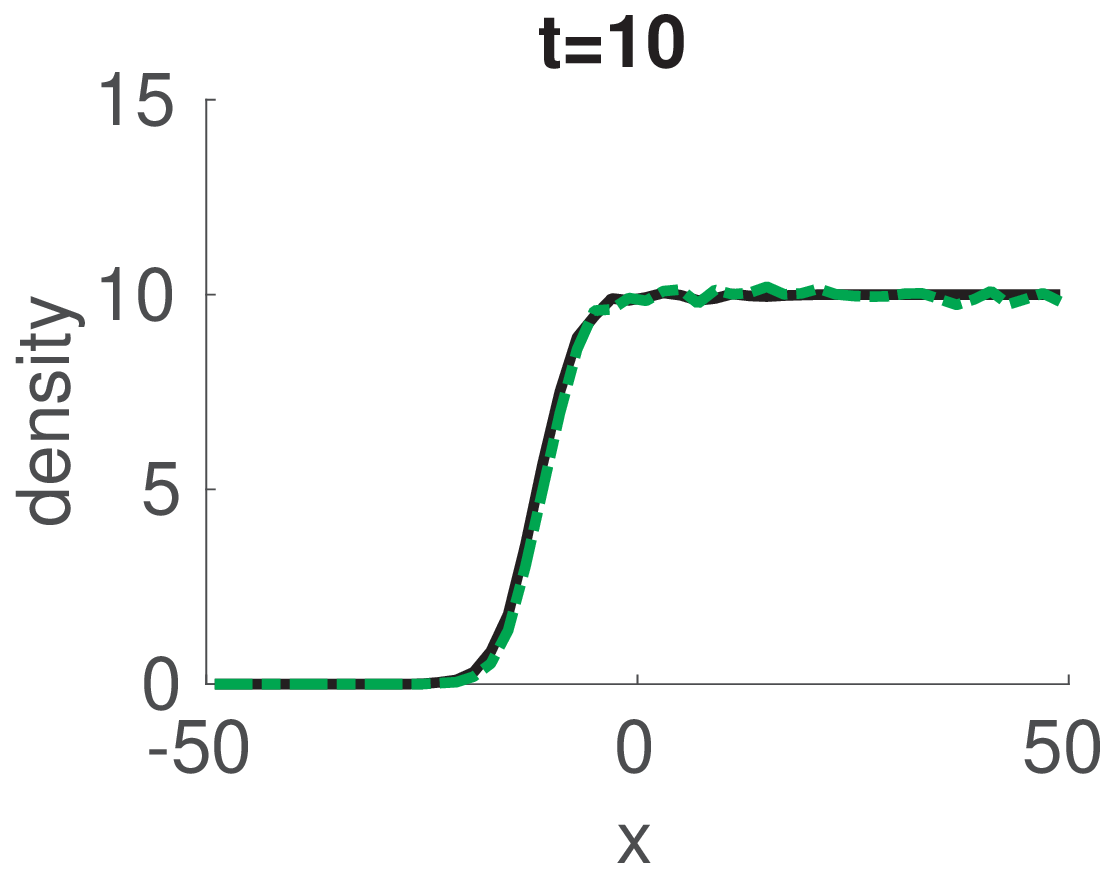}
\label{figure:tw_moving_60_algorithm_1}
}
\subfigure[]{
\includegraphics[width=0.31\columnwidth]{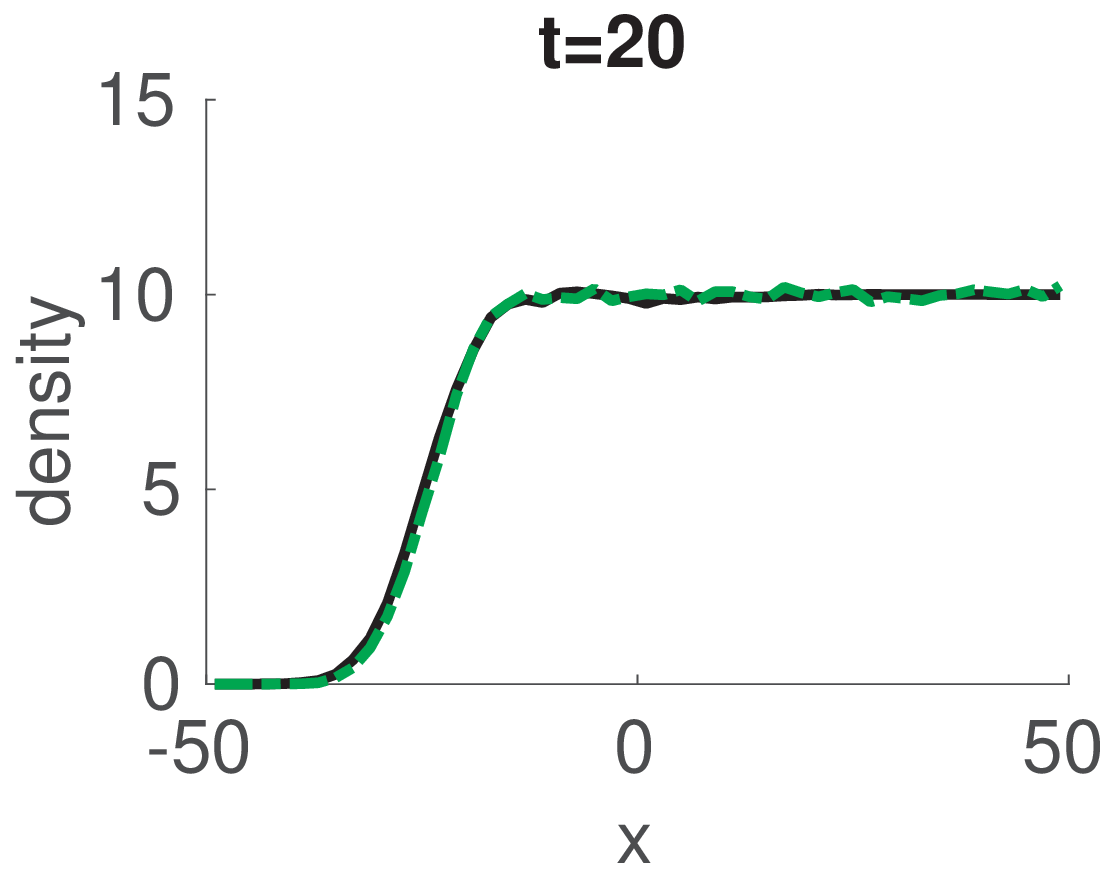}
\label{figure:tw_moving_60_algorithm_2}
}
\end{center}
\caption{Simulating a travelling wave using the adaptive hybrid algorithm.  
Parameters used are as described in Figure \ref{figure:tw}, with $u_{max}=10.5, \delta u=1.0, \eta =50$ for adaptive movements of the interface.
The black line shows the results of the adaptive hybrid algorithm, while the dashed green line shows the fully stochastic model. Panels \subref{figure:tw_moving_initial_condition},
\subref{figure:tw_moving_60_algorithm_1} and
\subref{figure:tw_moving_60_algorithm_2} show the particle density at times $t=0$, $t=10$ and $t=20$ respectively.}
\label{figmytwadaptivet10}
\end{figure}

In the particular case of the travelling wave considered in the previous section (\ref{section:travelling}), it is important that we keep the entire front of the wave in $\Omega_c$, since it is the description governing the wave front that dictates the wave speed. 
To ensure this, we take $u_{max} =10.5, \delta u=1.0$ for model parameters as in Figure \ref{figmytwadaptivet10}.

When we have performed several iterations of the hybrid adaptive algorithm and wish to take an average of the results we encounter some difficulties. 
After a full iteration of the algorithm has been completed, the interface between the models will have, in general, changed position following the wave front.
However, upon repeating the iteration, the position of the interface may have changed by a different amount.
This is due to the stochastic nature of the process that we are simulating.
%
We note that in the overlap region both of the model descriptions are valid.
With this in mind, we record the concentration in both the stochastic and deterministic regions for each iteration of the algorithm and combine the concentrations together to give an average value for the concentration at each position.
That is we take, for any point in the overlap region for any of the iterations of the algorithm, $u_{av}(x,t) = (A(x,t) + u_p(x,t))/2$, where $u_{av}$ is the concentration in the overlap region, $u_p$ is the concentration in the PDE-based region and $A$ is the particle number in the compartment-based region. 
Otherwise outside the regions covered by the overlap region, we use the deterministic and stochastic descriptions as usual.
It is this combination of deterministic and stochastic descriptions that is plotted in Figure \ref{figmytwadaptivet10}.

\noindent 
\begin{minipage}{\textwidth} 
  \centering 
\begin{center}
 \begin{tabular}{|p{12cm}|}  
\hline
  (i) Initialize and apply initial conditions. Set $t:=0$ and $k:=0$. \\
  (ii) If $k= \eta $, where $\eta$ is the checking interval, then check position of interface, otherwise proceed to step (iv) \\
  (iii) If $A_i > u_{max}$ for $i \in \{ -1, ..., -m \} $, and $u_{j}>u_{max}$ for $j \in \{ l-w,l+w \} $,
where the $l$th lattice point of the PDE lies on the interface $I_1$, and $w=\frac{h}{\Delta x_p}$ is the ratio of discretisations in $\Omega_c $ and $\Omega_p$, then update interface: $I_0:=I_0-h$.  \\
 If $A_i < u_{max}-\delta u$ for $i \in \{ -1, ..., -m \} $ and $u_{j}<u_{max}-\delta u$ for $j \in \{ l-w,l+w \} $, then update interface: $I_0:=I_0+h$.  \\
If $I_0$ has been updated, then density in newly created region is equal to density of that region in previous description.\\
(iv) Implement one iteration of Algorithm 1. 
Increment $k=k+1$.
Return to step (ii) unless final time is reached. \\
\hline 
\end{tabular}

\end{center}

  \medskip 

  Algorithm 2: Algorithm for stochastic reaction-diffusion simulations with an adaptive interface using a compartment-based region and a PDE-based region.  
\end{minipage}
$\vspace{6pt}$

Notable computational improvements are afforded by the hybrid model in comparison to the fully stochastic compartment-based model. Simulation time is decreased by a simulation-dependent factor of around 5. Note that the adaptive interface algorithm for the travelling wave simulations is significantly faster than the scenario with the fixed interface.

\begin{table}[h]
\begin{center}
\begin{tabular}{ | l | c | c | r |}
\hline			
  Model & Fully Stochastic Model (s) & Hybrid Model (s) & Speed up \\
\hline
  Simple diffusion (IC: mass in $[0,1]$) & 1381.5 & 260.6 & 5.3 x  \\
  Morphogen gradient & 2721.6  & 518.0 & 5.3 x \\
  Travelling wave (fixed interface) & 3133.3 & 688.1 & 4.6 x \\
  Travelling wave (adaptive interface) & 3133.3 & 527.6 & 5.9 x \\
\hline  
\end{tabular}
\end{center}
\caption{Computation times for each of the test problems, comparing the hybrid model with the fully stochastic model. Parameters used are as for Figures \ref{figure:RHS_gradient}, \ref{figure:morphogen_gradient}, \ref{figure:tw}, \ref{figmytwadaptivet10}. 
Speed ups are given as a multiple of the fully stochastic time.}
\end{table}

\subsection{Sensitivity analysis}\label{section:error_analysis}

\begin{figure}[h!]
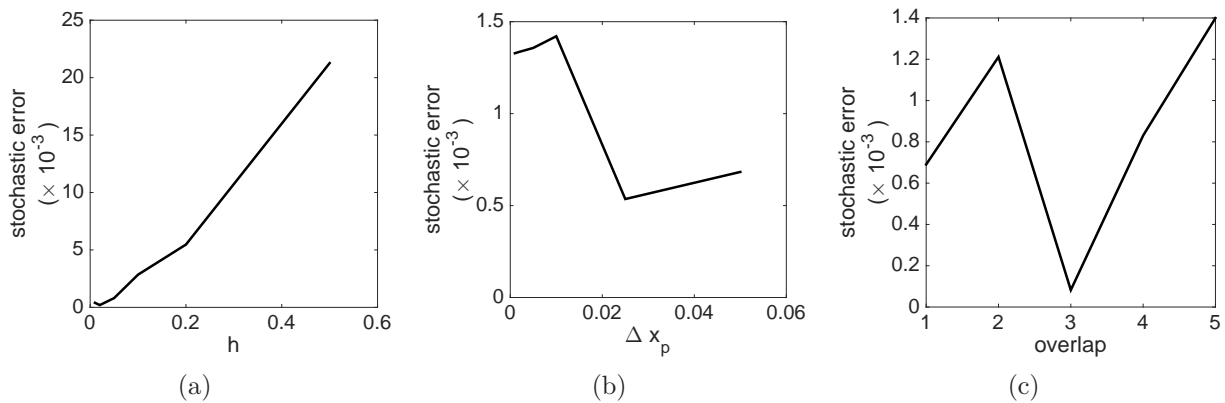

\begin{center}
\subfigure[]{
\includegraphics[width=0.31\columnwidth]{sensitivity_log_h2.eps}
\label{figure:sensitivity_h}
}
\subfigure[]{
\includegraphics[width=0.31\columnwidth]{sensitivity_log_dx2.eps}
\label{figure:sensitivity_dx}
}
\subfigure[]{
\includegraphics[width=0.31\columnwidth]{sensitivity_log_overlap2.eps}
\label{figure:sensitivity_overlap}
}
\end{center}
\caption{Sensitivity of the hybrid method to varying algorithm parameters: compartment size, $h$, PDE discretisation, $\Delta x_p$, and number of compartments in the overlap region, $m$. The stochastic error shown here is the absolute value of the difference between several repeats of a stochastic simulation and the analytic solution. Parameters used for simulations were as for Figure \ref{figure:LHS_gradient}, with a total of 1000 particles and 10 repeats. 
Panels \subref{figure:sensitivity_h},
\subref{figure:sensitivity_dx} and
\subref{figure:sensitivity_overlap} show the relative error for $h$, $\Delta x_p$, and $m$ respectively.}
\label{combinederrorplot}
\end{figure}

We demonstrate robustness of the coupling algorithm to choices of the algorithm parameters $h$, the compartment width, and $\Delta x_p$ the PDE discretisation, showing how the total error varies as a function of these parameters.
Since we are also able to vary the size of the overlap region in our coupling algorithm, we also demonstrate the effects of varying the number of compartments in this region.
As the test problem here, we use simple diffusion with the same step-function initial condition as in Figure \ref{figure:LHS_gradient} given by equation \eqref{RHS_IC}.
The results are presented in Figure \ref{combinederrorplot}.
The total error $E$ is calculated by summing the absolute value of the point-wise differences between the analytical and the hybrid solutions at the centre of each compartment in $\Omega_c$ equivalently in $\Omega_p$. 
The error is shown for a single time point, at $t=1$.

As $h$ increases, it is clear that the stochasticity in the error values increases due to the smaller number of compartments used. 
However, this is the behaviour we would expect and is also seen in the fully stochastic model.
With varying $\Delta x_p$, the magnitude of the stochastic error remains approximately constant.
Similarly, the error is independent to changes in the number of compartments in the overlap region.

\section{Discussion}\label{section:discussion}
\subsection*{Summary}
In this article, we have presented a novel hybrid algorithm for coupling a stochastic compartment-based model with a deterministic PDE model for reaction-diffusion systems.
This technique is helpful for simulating reaction-diffusion systems, providing most benefit in comparison with existing methods in cases where a detailed description is necessary in a part of the domain of interest, but there are computational restrictions preventing the use of the detailed stochastic model throughout the domain. 
We utilise an overlap region where both modelling descriptions are valid.
To perform the coupling, we apply a flux-based condition at one interface and a Dirichlet type condition at the other interface.
Furthermore, we justified mathematically the particular form of the boundary conditions used. 

Biochemical systems where reaction-diffusion modelling approaches have been applied are found widely in the natural world from population ecology \citep{murray2002mathematical}, to the spread of epidemics \citep{murray2002mathematical}, to cell biology such as calcium signalling \citep{rudiger2007hybrid}, and wound healing \citep{sherratt1990models}. 
In particular we focused on systems with multiple scales where detailed modelling is required in a certain region, but it might prove computationally wasteful to apply that method throughout the domain.
Such systems occur frequently in a biological context due to the multiscale nature of biological systems \citep{meier2009multiscale}.

The hybrid algorithm that we have developed was robustly tested and demonstrated by applying it to several biologically motivated problems in section \ref{section:numerical_simulations}. 
There are noteworthy improvements in simulation time in comparison to a fully stochastic model, including a decrease in simulation time by approximately a factor of 5 when applied to a suite of standard test problems. 
The performance of this hybrid algorithm and the error compared to an analytic solution was analysed and explained.

At low particle numbers, a deterministic modelling method may no longer be appropriate and a stochastic method should be applied to account for the variation.
There are disadvantages to the stochastic methods too; in particular they can require long simulation times.
In order to make best use of the complementary advantages of deterministic and stochastic models, multiscale hybrid models are becoming increasingly widespread, particularly in applications relating to reaction-diffusion systems.
We have presented our own hybrid coupling algorithm to segue between stochastic compartment-based models and deterministic PDE-based models. 
Further computational improvements have been reached by adding an adaptive interface to the algorithm.




%
%

\begin{acknowledgments}
J.U.H. would like to thank the London Mathematical Society/Nuffield Foundation for an Undergraduate Research Bursary.
\end{acknowledgments}

\bibliography{Harrison2014HAFCreferences.bib}        
\bibliographystyle{unsrtnat}
\end{document}